\documentclass[12pt]{iopart}
\usepackage[dvips]{graphicx}
\usepackage{amssymb}
\usepackage{amsfonts}

\begin{document}
\title[Beyond the Linearity of Current-Voltage Characteristics in Multiwalled Carbon Nanotubes]{Beyond the Linearity of Current-Voltage Characteristics in Multiwalled Carbon Nanotubes}

\author{B. Bourlon$^{1,2}$, C. Miko$^3$, L. Forr\'o$^3$, D.C. Glattli$^{1,4}$ and A. Bachtold$^{1,5}$}
\address{$^1$ Laboratoire Pierre Aigrain, Ecole Normale Sup\'{e}rieure, 24
rue Lhomond, 75231 Paris 05, France}
\address{$^2$ Department of Applied Physics, California Institute of Technology, Pasadena, California 91125, USA}
\address{$^3$ EPFL, CH-1015, Lausanne, Switzerland} \address{$^4$ SPEC, CEA Saclay,
F-91191 Gif-sur-Yvette, France}
\address{$^5$ ICN and CNM-CSIC, Campus Universitat Autonoma de Barcelona, E-08193 Bellaterra, Spain}

\begin{abstract}
We present local and non-local electron transport measurements on individual multi-wall nanotubes for bias voltage between 0 and about 4 V. Local current-voltage characteristics are quite linear. In contrast, non-local measurements are highly non-linear; the differential non-local conductance can even become negative in the high-bias regime. We discuss the relationship between these results and transport parameters such as the elastic length, the number of current carrying shells, and the number of conducting modes.

\end{abstract}
\submitto{\SST} \maketitle

\section{Introduction}
Multiwalled carbon nanotubes (MWNT) have recently attracted a lot
of attention as possible interconnects in future integrated
circuits $\cite{weiapl2001,kreume2002,liapl2003}$. MWNTs can carry
several hundreds of $\mu$A in a section of about 100 nm$^2$,
which corresponds to a current-carrying capacity much larger than
the one of today's interconnects. MWNTs consist of several nested
cylindrical graphene shells. Their electronic transport properties
have been studied mainly with devices where the electrodes are in
direct contact with the outer shell. Understanding the electronic
transport properties of a MWNT requires knowledge of
the transport properties of each shell. More precisely, the
conductance of a MWNT depends on parameters such as $l_e$ the mean
free path, $N_{shell}$ the number of current-carrying shells, and
$N_{mode}$ the number of conducting modes. It has been possible to
determine these parameters mainly in two distinct regimes. The two
regimes, which are set by the bias voltage applied to the
conductor, are the low-bias regime (voltage below $kT=25$ meV) and
the high-bias regime (around a few volts).

Results from previous studies on how
the mean free path, the number of shells, and the number of modes
per shell contributing to conduction vary between the low- and
the high-bias regime are reviewed in this paragraph. We note that these results have been
obtained on MWNT devices that have been fabricated following a
specific fabrication process (see below). For samples prepared using
a different method the results can be different
$\cite{a1,a2,a3,a3b,a4,a5,a6,a7}$. For example, the doping and the
mean-free path can be different due to a different device
preparation. Also, in some of these papers, the MWNT is suspended
so that the heat produced by Joule heating is less evacuated.

- $l_e$ the mean free path. In the low-bias regime, it has been
found that the transport is diffusive with a mean free path
$l_{e}$ $\lesssim$ $100\ nm$
$\cite{bachtoldnature1999,schAPA1999,bourlonprl2004n2}$. The
diffusion arises from disorder but the microscopic nature of these
scattering centers is not yet known. At high-bias, the mean-free
path becomes shorter with a mean free path $l_{e} \sim 5-10\ nm$
$\cite{yaoprl2000,javeyprl2004,bourlonprl2004n1}$. This regime has
been intensively studied these last few years for SWNTs. The reduction
of $l_{e}$  is attributed to the strong coupling between electrons
and optical phonons
$\cite{yaoprl2000,maulprl2004,dubayprb2003,ferrari}$.

-$N_{mode}$ the number of modes per shell participating
significantly in conduction. In the low-bias regime, it has been
shown that MWNTs are doped with a Fermi level that is hundreds
millivolts away from the charge neutrality point
$\cite{krugerAPL2001}$. The corresponding number of modes is about
$N_{mode}=10-30$. In the high-bias regime, the number of modes
that significantly contribute to conduction is suppressed to a
few modes $\cite{bourlonprl2004n1,collins2001,collinsprl2001}$.
This has been attributed to the Zener backscattering that occurs
between the lower and upper non crossing subands
$\cite{bourlonprl2004n1,ananprb2000}$.

-$N_{shell}$ the number of shells participating significantly in
conduction. In the low-bias regime, non-local resistance
measurements showed that essentially two shells contribute to
conduction when the distance between electrodes is below around $1
\mu m$ $\cite{bourlonprl2004n2}$. This changes at high bias where
most of the shells carry current. This result has been obtained
from the electrical breakdown experiments, which consist of
removing the shells one by one by injecting a large current in the
MWNT $\cite{a4,bourlonprl2004n1,collins2001,collinsprl2001}$.

Between the low- and high-bias regimes there are no experimental
methods to probe $l_e$, $N_{mode}$, and $N_{shell}$. The purpose
of this article is to discuss the relationship between the
current-voltage (I-V) characteristics of MWNTs and the variations
of $l_e$, $N_{mode}$, and $N_{shell}$ with bias voltage.
Surprisingly, the local I-V characteristics appear mostly linear
up to several volts. At first sight, this may suggest that $l_e$,
$N_{mode}$, and $N_{shell}$ do not change as the bias voltage is
increased, in opposition to the above summary. However, we will
see that the variation of $l_e$, $N_{mode}$, and $N_{shell}$ can
be observed in non-local I-V characteristics that are highly
non-linear.

\section{Experimental results}

The MWNTs are synthesized by arc-discharge evaporation and
carefully purified $\cite{bonaram1997}$. Nanotubes are sonicated in dichloroethane and
then dispersed on an oxidized Si substrate. Finally, they are
contacted by top Cr/Au electrodes using electron-beam lithography.
Typical two-point resistances range from 5 to 30 $k\Omega$.

We carry out local and non-local measurements. In non-local
measurements, the voltage drop is measured outside the region
between the current biased electrodes, while for local
measurements the current that flows through the voltage biased
electrodes is measured (see schematic in Fig. 1(b)) Local
measurements in Fig. 1(a) show that the current is roughly linear
with the bias voltage up to 3 volts. This could suggest that
parameters such as $l_e$, $N_{mode}$, and $N_{shell}$ remain
constant with the bias voltage. However, Fig. 1(b) shows that the
non-local measurements are highly non-linear. The non-local
voltage $\Delta V$ first increases linearly up to 1.5 V and then
decreases. Similar results are obtained for several other devices.
This is the main result of the paper. It shows that the transport
characteristics of MWNTs change between the low- and high-bias
regimes.

It was shown that the non-local voltage at low-bias mainly results
from the current pathway through the two outermost shells
$\cite{bourlonprl2004n2}$.  The current pathway at low bias was
shown in Ref. $\cite{bourlonprl2004n2}$ to be described by a
resistive transmission line with the intrashell resistance about
$\sim$ 10 k$\Omega$/$\mu$m and the intershell conductance about
$\sim$(10 k$\Omega$)$^{-1}$/$\mu$m, see Fig. 4(a). In Fig. 1(b)
the reduction of $\Delta V$ at high bias suggests that the current
pathway in the MWNT changes. Less current flows through the
outermost shell in regions that are not between the biased
electrodes. A change in the current pathway between the low- and
the high-bias regimes is not surprising when considering the
variations of $l_e$, $N_{mode}$, and $N_{shell}$ discussed in the
introduction.

When looking at the high non-linearity of non-local measurements,
it is however surprising that the local $I-V$ curves appear so
 linear. To try to clarify this issue, we have studied more
than 50 different nanotubes. Three representative measurements of
local $dI/dV-V$ characteristics are presented in Fig. 2. Between 0
and about 2 V, we obtain all the possible slopes for $dI/dV-V$;
$dI/dV-V$ can increase, remain constant, or decrease. At higher
voltages, however, $dI/dV$ systematically decreases. The different
variations of $dI/dV-V$ suggest that transport characteristics,
such as $l_e$, $N_{mode}$, and $N_{shell}$, vary differently for
different tubes as the voltage is increased.

Figure 3 summarises the local measurement over all the tubes. The figure shows that there is a correlation between the slope of $dI/dV-V$ and
geometrical parameters of MWNTs such as the length and diameter. A reduction of $dI/dV-V$ is more probable to occur for
long and thin MWNTs, while an enhancement is more probable for short and thick tubes.

It is interesting to compare these results to measurements on
SWNTs. The enhancement of $dI/dV-V$ can be observed at low bias
voltage up to about 100 mV, which is usually attributed to Coulomb
interaction $\cite{bockrath1999}$. It is however very rare to
observe in SWNTs an enhancement of $dI/dV-V$ up to 2V as it is the
case for short and thick MWNTs. This difference can be attributed
to the large diameter shells in MWNTs that have shorter separations
in energy between subbands. High energy subbands are then easier
to populate with electrons and thus can easily carry current at high bias
voltage. In addition, the different behaviour between SWNTs and
short thick MWNTs can also be attributed to the fact that MWNTs
consist of several SWNTs. The enhancement of $dI/dV-V$ can reflect
the enhancement of the number of shells that carry current.
Further discussion will be found in the section below.

\section{Discussion}

We now discuss the relationship between these results and transport parameters such as $l_e$, $N_{mode}$, and $N_{shell}$. Unfortunately, we are not at the stage where we are able to quantify local and non-local $I-V$ characteristics in MWNTs. We can neither account for the length nor the diameter dependences of the local $I-V$ characteristics reported in Fig. 3. However, in the following, we will explain our understanding of how $l_e$, $N_{mode}$, and $N_{shell}$ change from the low- to the high-bias regime.

In Fig. 4(a), we describe MWNTs by multiple resistive lines. Each shell is modeled by a
series of resistances that depend on both the number of
modes and the transmissions of these modes. This model has successfully described measurements in the low-bias regime in which the current is shown to flow through the two outermost shells $\cite{bourlonprl2004n2}$. As the bias voltage is increased, more shells are expected to participate in the conduction. Moreover, the number of modes per shell, and
the transmission of each of these modes are also expected to change. The main problem is to
know how these parameters change with the bias voltage.

We first look at the transmission $T(V)$. Yao et al.
$\cite{yaoprl2000}$ have shown that the current in SWNTs saturates
at high voltage. This has been attributed to the suppression of
the transmission due to scattering processes between electrons and
optical phonons with an energy of $\hbar \Omega = 160$ $meV$. At
first sight we can assume that an electron is scattered as soon as
its energy acquired from the electric field reaches $\hbar \Omega$
so that the mean free path $l$ is given by $e(V/L)l=\hbar \Omega$.
The transmission, which is obtained by dividing $l$ by $L$, is
then $T(V)=\hbar\Omega/eV$. The conductance is obtained using the
Landauer formula (Fig. 4(b)). The figure shows how the
differential conductance goes to zero as the bias voltage is
increased.

We now discuss the variation in the number of modes that
participate in conduction as the bias voltage is increased. For
simplicity, we start the discussion for undoped tubes. As a first
approximation, one may simply count the number of subbands that
lie between the energy range $eV$. In the high-bias regime there
would be several tens of such subbands for a 10 nm diameter shell.
This however overestimates the number of efficient modes that
contribute significantly to the current. It has been found that
all shells carry a very similar current at high-bias and that this
current is close to the saturation current of about 20 $\mu A$
which is flowing through metal SWNTs. This suggests that the
number of efficient subbands per shell in MWNTs is close to the
number of conducting modes in SWNTs, which is 2.

More precisely, this reduction of efficient modes has been
attributed to intersubband Zener tunnelling. This process involves
electrons tunnelling from the top of valence subbands to the
bottom of conduction subbands (see Fig. 5(a)) $\cite{ananprb2005}$. The transmission of
this process is dramatically lower for electrons that lie in
subbands far in energy from the charge neutrality point, so that
such subbands do not contribute much to the conduction.

Zener tunnelling is expected to be voltage dependent. The
transmission is expected to become higher as the bias voltage is
enhanced, since the electric field reduces the length of the
barrier. In addition, the high current in the MWNT enhances the
temperature through Joule heating $\cite{chen2005}$ $\footnote{In
Ref. $\cite{chen2005}$ the temperature has been shown to reach
about 2000 C in the high-bias regime. These measurements have been
carried on suspended MWNTs. In our case, the MWNTs are lying
directly on the silicon oxide layer, so that the produced heat can
be better evacuated. It makes difficult to estimate the
temperature in our experiments; it can be between 300 and 2000
K.}$, so that electrons can easier pass the Zener barrier. This
results in an increase of the differential conductance as shown
in Fig. 4(c).

We now take into account the fact that our tubes are significantly
p-doped $\footnote{This can be shown by measuring the conductance
as a function of the voltage applied on the backgate. The
conductance is observed to be reduced by around a factor 2 as the
gate voltage is swept from -100V to +100V. We never observed an
enhancement of the conductance as the gate voltage is increased
even at very large positive gate voltages.}$. Kruger et al. have
shown using an electrochemical gate that the Fermi level is
shifted by around 0.3 eV from the charge neutrality point, which
corresponds to $N_{mode} =10-30$ $\cite{krugerAPL2001}$. Figure
5(b) shows the schematic of a semiconducting shell with a few
subbands since it is difficult to represent much more subbands.
The number of conducting modes is here $N_{mode} =4$ and the
states contributing to transport are represented by the dark gray region
set by $eV$. Figure 5(c) shows the case at high-bias. We assume
that electrons travel at constant energy between the electrodes.
Most electrons have then to cross the gap via Zener tunneling.
Only electrons in subbands close to the charge neutrality point can then
carry current since the others are backscattered when they try to
cross the gap. Thus, many of the subbands that conduct at low-bias
become inefficient at high-voltage. We notice however that these
subbands generate a low but non-negligible current since some
electrons do not have to cross the gap, represented in the dark gray region in Fig.
5(c). Overall, as the bias voltage is increased from zero, the
number of modes $N_{mode}$ that carry a significant amount of current first
decreases due to the Zener backscattering at the gap. At higher
voltage, however, the differential conductance increases again
since the Zener transmission is enhanced.

We here turn our attention to the number of shells that contribute
significantly to electron transport. Adding conducting shells
obviously increases the differential conductance. It has been
shown that in the linear regime only one or two shells participate
in the conduction $\cite{bourlonprl2004n2}$. This is different
from the high-bias regime where almost all shells have been shown
to participate
$\cite{a4,bourlonprl2004n1,collins2001,collinsprl2001}$. However
little is known about how $N_{shell}$ increases between the linear
and high-bias regime (Fig. 4(d)). Indeed, the probability for an
electron to go into the adjacent shell depends on the shell
resistances that are given by $l_e$ and $N_{mode}$
$\cite{trioprb2004,c1,c2}$, as illustrated in the resistive line
model in Fig. 4(a). The probability is enhanced (reduced) as the
resistances are larger (smaller). Because the behaviour of
$N_{mode}$ is not well understood, it is difficult to predict how
electrons penetrate in inner shells as the voltage increases.

The variation of $dI/dV-V$ can also be a consequence of the
contact between the nanotube and the electrode. Indeed, it has
been shown that $dI/dV-V$ increases with $V$ due to Coulomb
interaction. When the resistance of the contact is larger than $
\sim h/2e^2$, the functional form is expected to follow a power law
$\cite{b1,b2,b3,b4}$. The exponent has been measured to be about 0.3.
For lower contact resistance, the functional form of $dI/dV-V$
remains to be determined. However, the enhancement of $dI/dV-V$ is
expected to be much less pronounced. It is thus suitable to work
with low-ohmic contacted MWNTs in order to disregard the
contribution from the contacts in the analysis of $I-V$
characteristics.

\section{Conclusion}

We have presented local and nonlocal measurements in MWNTs. The
main result is that nonlocal $I-V$ characteristics are highly non
linear. The nonlocal voltage can even decrease at high bias
voltage. This suggests that the current pathway changes as the
bias voltage increases. This is attributed to the variation of
transport parameters such as $l_e$, $N_{mode}$, and $N_{shell}$.
As the bias voltage increases, the mean-free path is expected to
decrease, while the number of shells is expected to increase. The
behaviour of the number of modes that significantly contributes to
transport is expected to be more complicated when working with
doped MWNTs; $N_{mode}$ first decreases and then increases.
Further studies are needed to understand transport properties of
MWNTs between the low and high-bias regimes. In particular, it
would be interesting to account for the length and diameter
dependences of the $I-V$ characteristics that are reported in Fig.
3. For this, it would be very interesting to study DWNTs, which are
simpler systems.

\ack
We thank B. Pla\c{c}ais for discussions, and P. Morfin, F. R. Ladan and
C. Delalande for support. LPA is CNRS-UMR8551 associated to University Paris 6 and 7.
The research has been supported by the DGA, ACN, sesame, the Swiss National Science
Foundation, and its NCCR "Nanoscale Science".

\section*{References}
\bibliographystyle{prsty}
\bibliography{./biblio}

\clearpage
\begin{figure}[htp]
\centering
\includegraphics[width=10cm,keepaspectratio,clip]{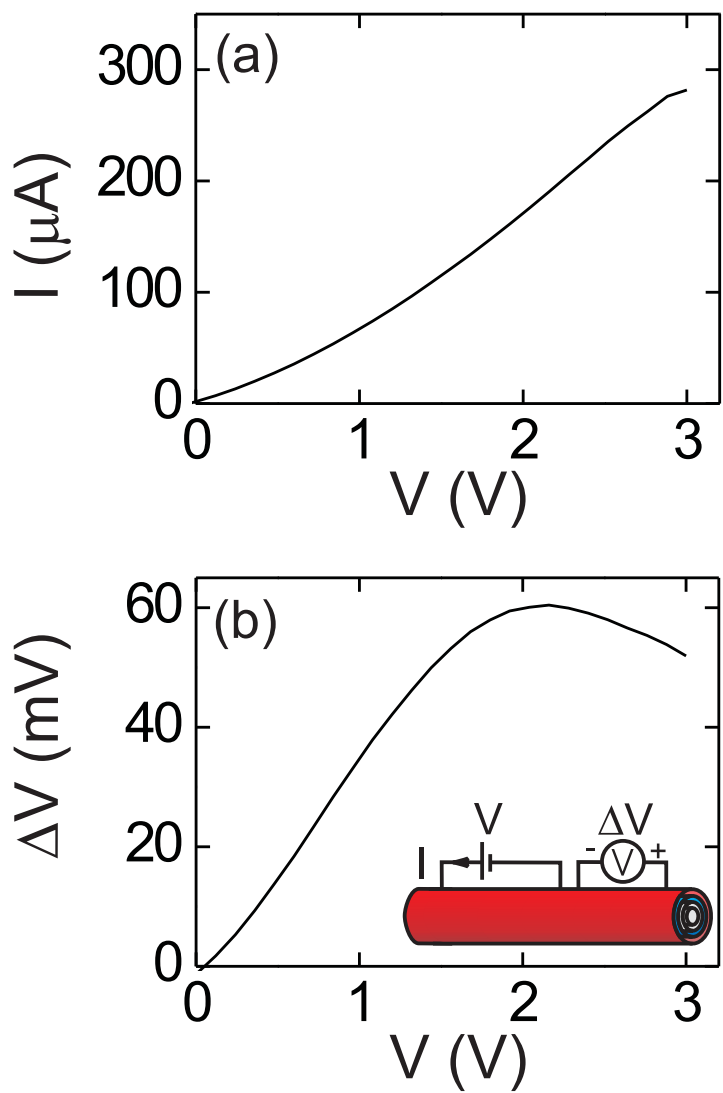}
\begin{center}
\caption{\label{fig1.fig} \footnotesize{(a) Local and (b)
Non-local measurements on a same MWNT at room temperature in air.
The schematic shows the measurement setup. The non-local
measurements are obtained with four electrodes that contact the
MWNT, while the local measurements are obtained with two
electrodes. The voltage $V$ is applied, while the current $I$ is
recorded for the local measurement and $\Delta V$ is recorded for
the non-local measurement}}
\end{center}
\end{figure}
\clearpage

\clearpage
\begin{figure}[htp]
\centering
\includegraphics[width=12cm,keepaspectratio,clip]{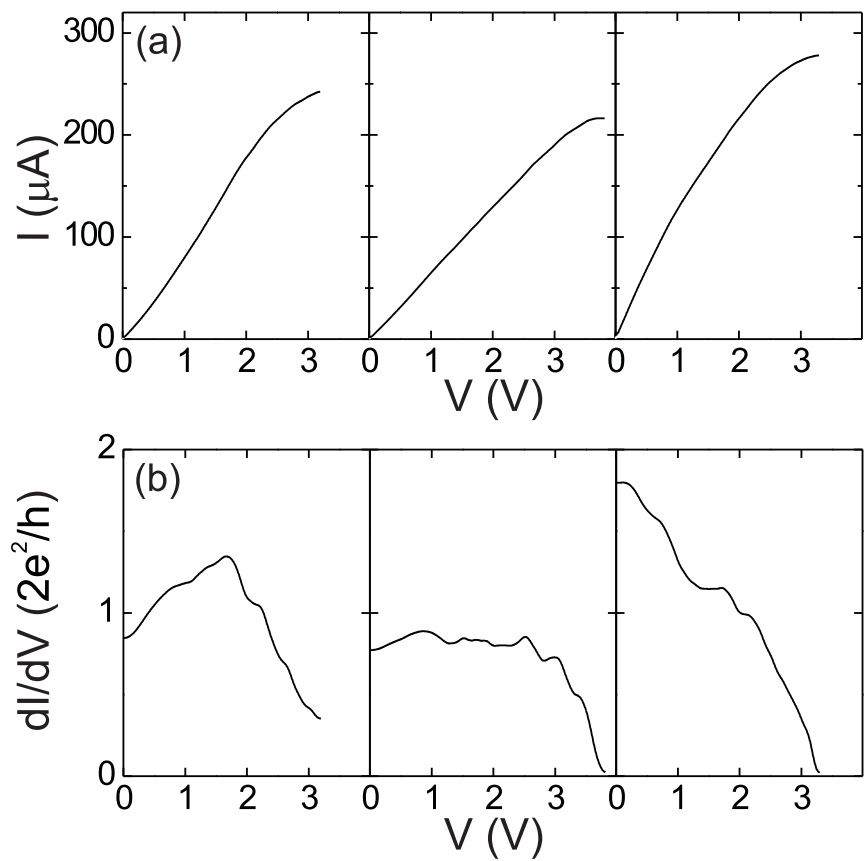}
\caption{\label{fig2.fig}
\footnotesize{(a) Local current-voltage
characteristics for different MWNT devices. (b) Corresponding
differential conductances. All the possible slopes for $dI/dV-V$
can be obtained below about 2 V; $dI/dV-V$ can increase, remain
constant, or decrease.}}
\end{figure}
\clearpage

\clearpage
\begin{figure}[htp]
\centering
\includegraphics[width=12cm,keepaspectratio,clip]{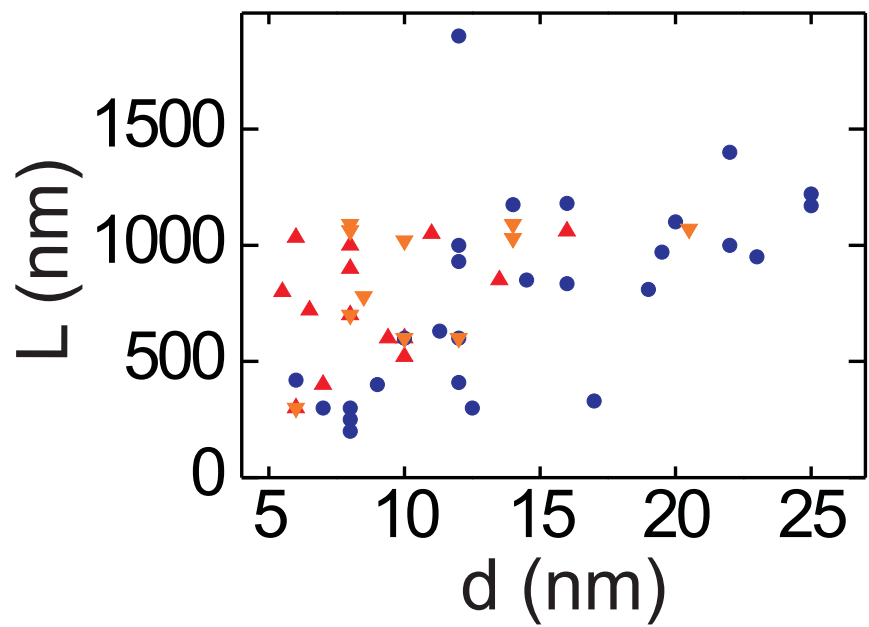}
\caption{\label{fig3.fig} \footnotesize{Slope of the local
differential conductance (V below about 2V) for more than 50 MWNT
devices as a function of diameter and length. The blue discs
correspond to positive slopes, the orange triangles pointing
downwards to constant slopes, and the red triangles pointing
upwards to negative slopes.}}
\end{figure}
\clearpage

\clearpage
\begin{figure}[htp]
\centering
\includegraphics[width=14cm,keepaspectratio,clip]{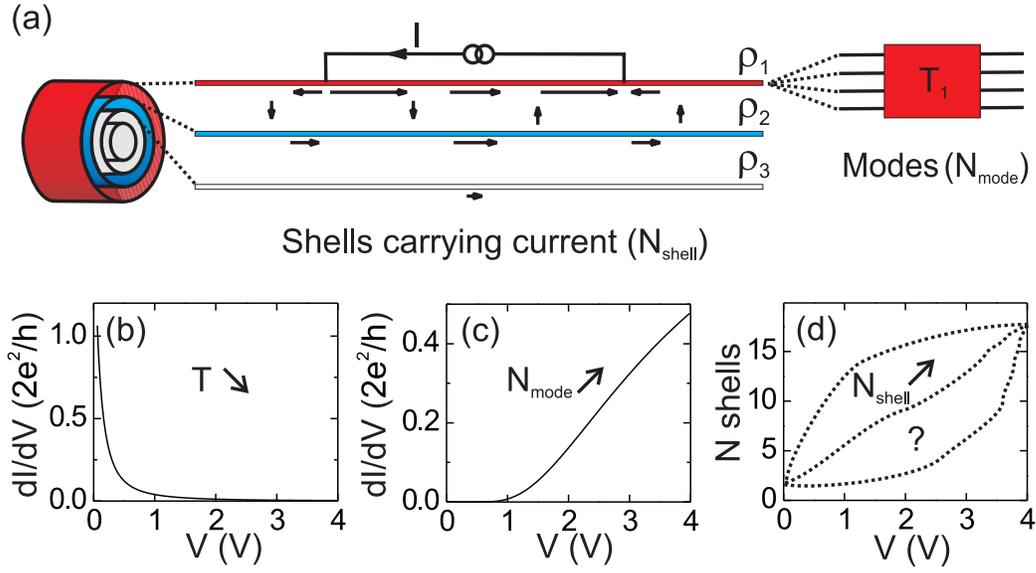}
\caption{\label{fig4.fig} \footnotesize{(a) Schematic of a
multiple transmission line that describes transport of MWNTs. (b)
The calculated contribution of $dI/dV-V$ that results from the
scattering between electrons and optical phonons. (c) The
calculated contribution of $dI/dV-V$ that results from the Zener
tunneling. The tube is undoped and semiconducting. The enhancement
of $dI/dV-V$ corresponds to an enhancement of modes that
contribute significantly to transport. The differential
conductance is obtained from the Zener transmission
$T_{Z}=\exp(-\frac{4\sqrt{2m^*}LE^{3/2}}{3e\hbar V})$ with a shell
length $L=1$ $\mu m$, an effective mass $m^{*}=6.5 \times
10^{-33}kg$ and a gap $E=53$ $meV$ $\cite{bourlonprl2004n1}$. The
values of $m^{*}$ and $E$ correspond to a diameter of $15$ $nm$.
(d) The number of current-carrying shells (and so $dI/dV$)
increases with the bias voltage. The behavior of $N_{shell}$ as a
function of $V$ is not known. The curves are drawn by hand.}}
\end{figure}
\clearpage

\clearpage
\begin{figure}[htp]
\centering
\includegraphics[width=8cm,keepaspectratio,clip]{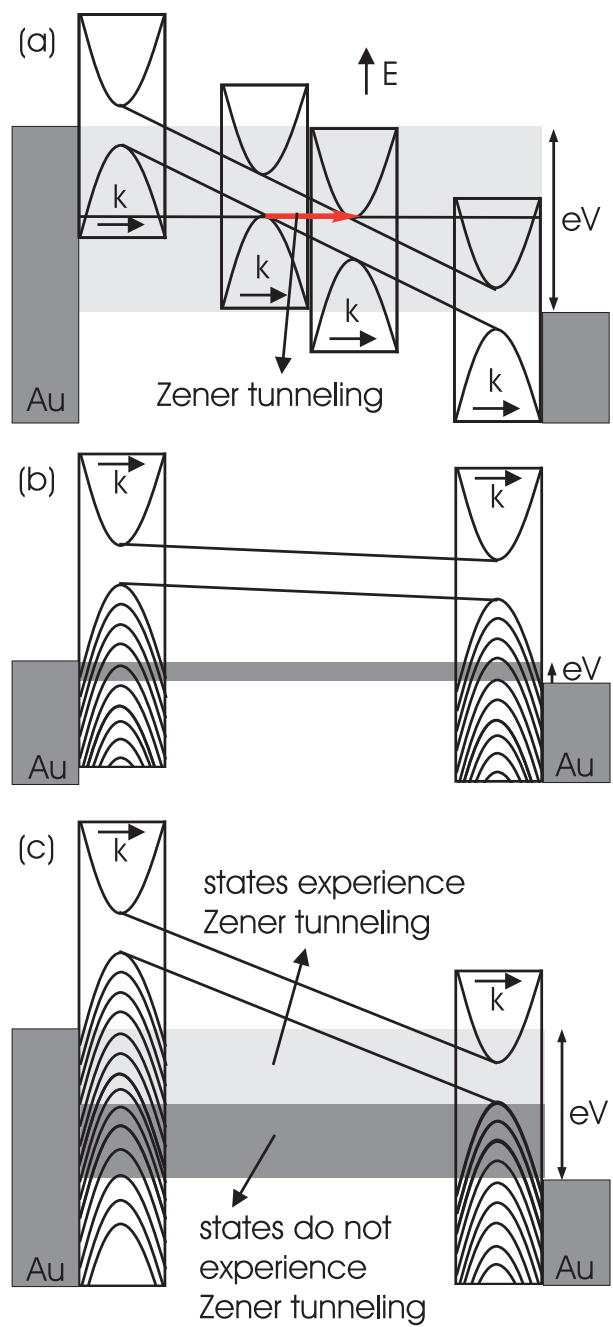}
\caption{\label{fig5.fig} \footnotesize{Schematic of the potential
variation in space. The boxes show the band diagram of a
semiconducting shell. (a) Zener tunneling from the top of the
valence subband to the bottom of the conduction subband. The shell
is undoped. The linear voltage drop along the tube results from
the short mean-free path that arises from the scattering between
electrons and optical phonons. (b) Low-bias regime for a doped
shell. (c) High-bias regime for a doped shell.}}
\end{figure}
\clearpage

\end{document}